\documentclass[fleqn,10pt]{wlscirep}
\newtheorem{theorem}{Theorem}
\newtheorem{proposition}[theorem]{Proposition}

\title{Measures of multisensory integration based on dependent probability summation: from spike counts to reaction times}

\author[1,*]{Hans Colonius}
\author[2]{Adele Diederich}
\affil[1]{Carl von Ossietzky Universit\"{a}t Oldenburg, Department of Psychology, Oldenburg, 26111, Germany}
\affil[2]{Jacobs University Bremen, Psychology and Methods, Bremen, 28759, Germany}

\affil[*]{hans.colonius@uol.de}

\begin{abstract}
A single neuron is categorized as``multisensory'' if there is a statistically significant difference between the response evoked by an audio-visual stimulus combination and that evoked by the most effective of its components individually. Crossmodal enhancement is commonly expressed as a proportion of the strongest unisensory response. However, being responsive to multiple sensory modalities does not guarantee that a neuron has actually engaged in integrating its multiple sensory inputs, rather than simply responding to the most salient stimulus. Here, we propose an alternative index measuring by how much the crossmodal response surpasses the level obtainable by optimally combining the unisensory responses. Optimality is defined by probability summation  combining the unisensory responses under maximal negative stochastic dependence. The new index is analogous to measuring crossmodal enhancement by the amount of violation of the ``race model inequality'', which is widely used in reaction time studies of multisensory integration. Neurons previously labeled as ``multisensory'' may lose that property since the new index tends to be smaller than the traditional one. This is exemplified with a data set collected from single SC neurons. The new easy-to-compute index does not require any specific distributional assumption. It is sensitive to the variability in the data, in contrast to the traditional index which, by definition, only depends on the means of the uni- and crossmodal response distributions.
\end{abstract}
\begin{document}

\flushbottom
\maketitle
%
%
\thispagestyle{empty}


\section*{Introduction}
Single neurons in the deep layers of the mammalian superior colliculus (SC)  integrate afferent visual, auditory, and somatosensory cues and generate efferent motor commands to structures innervating the musculature of, e.g., the eyes and hands \cite{stein2012,steinmeredith93}. \textit{Multisensory integration} is defined operationally as the neural process by which  unisensory signals are combined to produce a multisensory response that is significantly different from the responses evoked by the modality-specific component stimuli \cite{Stein2010}. For example, at the level of a single superior colliculus (SC) neuron, response strength has traditionally been measured as the absolute number of impulses (spikes) registered within a fixed time interval after stimulus presentation or, sometimes, by the firing rate within this interval. A neuron is categorized as being  ``multisensory'' if the absolute number of spikes to a cross-modal, e.g. visual-acoustic, stimulus combination is significantly higher (or, in case of inhibition, lower) than the number of spikes evoked by the most effective of its components individually \cite{steinmeredith93}. Moreover, if a neuron responds, for example, to visual but not to auditory stimulation and if the response to a visual-auditory combination differs significantly from the response to the visual stimulus, it is also considered being ``multisensory''.


%

Up to date, the most widely used descriptive measure of the magnitude of multisensory integration is the \textit{crossmodal enhancement index} ($\mathrm{CRE}$), also termed \textit{crossmodal interaction index}. It is defined as 
\begin{equation}\label{cre}
\mathrm{CRE}=\frac{\mathrm{CM}-\mathrm{SM_{max}}}{\mathrm{SM_{max}}}\times 100,
\end{equation}
where $\mathrm{CM}$ is the mean number of spikes in response to the crossmodal stimulus and $\mathrm{SM_{max}}$ is the mean number of spikes to the most effective modality-specific component stimulus \cite{meredithstein83}. Thus, $\mathrm{CRE}$ expresses crossmodal enhancement as a proportion of the strongest unisensory response. Some modifications of $\mathrm{CRE}$ have been proposed as well\cite{perrault2005}. Prominently, in the ``additive model'',  term $\mathrm{SM_{max}}$ in Equation~(\ref{cre}) is replaced by the sum of the unisensory responses \cite{populin2002}. The additive version has raised some controversy because, under some modeling assumptions, an additive combination of cross-modal inputs yields a prediction of optimal multisensory integration\cite{Pouget2002}. Thus, observing that a neural circuit is actually engaged in optimal multisensory enhancement but does not achieve ``superadditivity'', would lead one to conclude that no multisensory integration has taken place. Similarly, any crossmodal response larger than the largest unisensory response but smaller than the sum might be misinterpreted as response depression \cite{steinetal2009}. In summary, the issue of exactly how to measure the amount of multisensory integration has been under debate for some time \cite{steinetal2009,perrault2005,Beauchamp2005}.

The purpose of this paper is to suggest an alternative to existing measures of crossmodal response enhancement. While having descriptive value, the main weakness of $\mathrm{CRE}$ and related measures is that they lack a commonly accepted theoretical basis. Such a basis is essential since merely being responsive to multiple sensory modalities does not guarantee that a neuron has actually engaged in integrating its multiple sensory inputs, rather than simply responding to the most salient stimulus modality. As Stein and colleagues \cite{steinetal2009} (ibid, p. 114) have put it, ``\textit{At the time of the early physiology studies in the 1980s, it was considered possible that these neurons only represented a common route by which independent inputs from a variety of senses could gain access to the same motor apparatus in generating behavior (e.g., possibly employing a ``winner-take-all'' algorithm}).'' 

Given that the actual computations performed by a multisensory neuron are still not fully understood\cite{miller2015}, developing a new measure should not depend on specific assumptions about the multisensory integration process. The suggestion offered here is a measure that compares the mean observed cross-modal response of a neuron with the largest mean achievable by optimally combining its unisensory responses, but without actually integrating them. Specifically, it measures by how much a neuron integrates information above the level obtainable by an optimal ``winner-take-all'' algorithm, as mentioned in the above quote. Note that the new measure does not presuppose that a neuron follow a specific operational mode. Rather, it takes the result of a potential probability summation mechanism as a benchmark to define the maximal enhancement that can be predicted by separately combining unisensory information streams. Because this  measure generally is  more restrictive than the traditional $\mathrm{CRE}$, many neurons previously categorized as ``multisensory'' risk losing that property. 

In order to motivate the new definition, we first consider an established measure of crossmodal enhancement in behavioral data, the race model inequality for reaction times. A numerical measure derived from that inequality turns out to be completely analogous to the measure proposed here for neural data. After introducing the new index, its properties are illustrated on a sample of spike data (Mark Wallace, personal communication, July 18, 2015). and compared to the traditional index. Finally, the special case of Poisson distributed spikes serves to demonstrate that, in contrast to the traditional index, the new one takes the variability of the data into account as well.
\section*{Measuring crossmodal enhancement of reaction time}
In the \textit{redundant signals paradigm}, stimuli from two (or more) different modalities are presented more or less simultaneously, and participants are instructed to respond to a stimulus of any modality, whichever is detected first. Besides comparing relative detection frequencies of unimodal vs. crossmodal stimuli, behavioral response strength is most often measured by reaction time (RT), that is, the time it takes a participant to respond (e.g., via button press) to a suddenly appearing stimulus, often visual or acoustic. Typically, time to respond in the cross-modal condition is shorter than that in either of the unimodal conditions. A significant reduction of mean RT to the cross-modal stimulus, compared with the faster of the unimodal mean RTs, has been taken as evidence for some true multisensory processing (``coactivation'') underlying the cross-modal reaction times\cite{miller82}. In analogy to $\mathrm{CRE}$ at the neural level, the index of \textit{crossmodal response enhancement for reaction time} ($\mathrm{CRE_{RT}}$) is defined as\cite{Cappe2009,stoep2015,diederichcolPP2004,Buchholz2012}
\begin{equation}\label{msirt}
\mathrm{CRE_{RT}}=\frac{RT_{\min}-RT_{CM}}{RT_{\min}}\times  100,
\end{equation}
where $RT_{CM}$ is the mean RT to the cross-modal stimulus and $RT_{\min}$ is the faster of the unimodal mean RTs. Thus, $\mathrm{CRE_{RT}}$ expresses multisensory enhancement as a proportional reduction of the faster unisensory response by the cross-modal response. For concreteness, we rewrite $\mathrm{CRE_{RT}}$ for the case of visual-auditory stimulation, with $\mathrm{E}RT_V, \mathrm{E}RT_A$, and $\mathrm{E}RT_{VA}$ denoting expected (mean) reaction time to the visual stimulus, the auditory stimulus, or  the visual-auditory stimulus combination, respectively. $\mathrm{CRE_{RT}}$ then becomes
\begin{equation}\label{crert}
\mathrm{CRE_{RT}}=\frac{\min\{\mathrm{E}RT_V,\mathrm{E}RT_A\}-\mathrm{E}RT_{VA}}{\min\{\mathrm{E}RT_V,\mathrm{E}RT_A\}}\times 100,
\end{equation}
Just as neural measure $\mathrm{CRE}$ of Equation~(\ref{cre}), index $\mathrm{CRE_{RT}}$ has descriptive value. For example, $\mathrm{CRE_{RT}}=10$  means that response to the visual-auditory stimulus is 10~\% faster than the faster of the mean responses to unimodal visual and auditory stimuli. 

\subsection*{The race model}
Interestingly,  it has been recognized early on\cite{raab62} that simply comparing crossmodal and unimodal mean RTs is not diagnostic with respect to a presumed underlying multisensory integration process, for the following reason. Let us assume that in the crossmodal condition, (i)~each individual stimulus elicits a process performed in parallel to the others and, (ii),~the finishing time of the faster process determines the observed RT. In this so-called race model, no actual integration of the unimodal processes takes place but, nevertheless, mean RT in the crossmodal condition is predicted to be shorter than the faster of the unimodal mean RTs, due to ``statistical facilitation'' (aka ``probability summation'' or ``winner-take-all'' mechanism). In order to gauge whether observed crossmodal RTs are faster than predicted by statistical facilitation,  Jeff Miller\cite{miller82,miller2016} proposed the race model inequality (RMI) test,
\begin{equation}\label{rmi}
P(\min\{V,A\}\le t) \le P(V\le t)+ P(A\le t) \;\;\mbox{or,}  \;\; F_{VA}(t) \le F_V(t) + F_A(t) \;\;\mbox{ for all }t, \; t \ge 0.
\end{equation} 
Here $V$ and $A$ denote visual and auditory processing times, respectively, with $F_V, F_A$ the corresponding unimodal RT distributions, and $F_{VA}$ the distribution of the RTs in the crossmodal (visual-auditory) condition, ignoring possible other stages, like response preparation, of observable RT. Violation of Equation~(\ref{rmi}) at any time point $t$ is evidence in favor of some form of multisensory integration taking place above statistical facilitation, often termed ``coactivation''. Note that  stochastic independence between the processing times $V$ and $A$ is not required, but the test is valid only if an assumption of ``context independence'' holds: the distributions of $V$ and $A$ in the unimodal conditions must equal their corresponding marginal distributions in the crossmodal condition\cite{luce86, colonius1990possibly} (see next subsection). 

The race model inequality has become the standard tool for testing whether observed reaction times to crossmodal stimuli are faster than predicted by a simple statistical facilitation mechanism. Gondan and Minakata\cite{gondan2016} report 83 studies from 2011 to 2014 performing the inequality test using a variety of statistical methods. Because, unlike $\mathrm{CRE}$, Inequality~(\ref{rmi}) does not represent a single numerical measure of the amount of crossmodal enhancement,  it has become practice to compute the following geometric measure: the area $S$ between $F_{VA}$ and $F_V + F_A$ defined by all $t$ values where the race model inequality is violated:
\begin{equation}
S=\int_0^\infty \mathbf{1}_C(t) \, dt \;\; \;\mbox{ with } \; C=\{t\,:\, F_{VA}(t) > \min\{F_V(t) + F_A(t),1\}.
\end{equation}
The sample estimate of area $S$ is then taken as index of the strength of violation of the inequality. Notably, a brief discussion of the race model inequality in the next section reveals that area $S$ can be interpreted as the expected value of random variable $\min\{V,A\}$ (under maximal negative dependence) and estimating $S$ is rather straightforward not requiring any geometric argument (for details, see also \cite{coldie06}). 
\subsection*{Context independence and coupling of random variables}
Sometimes, instead of Equation~(\ref{rmi}), a more restrictive inequality is tested,
\begin{equation}
   F_{VA}(t)  \le F_V(t)+F_A(t)- F_V(t)*F_A(t),\label{ind}
\end{equation}
assuming \textit{stochastic independence} between $V$ and $A$. This raises the general question of how the random variables in the unimodal conditions, $V$ and $A$,  related. Actually, as already observed by R.D. Luce \cite[p. 130]{luce86},  there exists --a-priori-- \textit{no stochastic relation} between them: the probability measures for $V$ and $A$, $P_V$ and $P_A$, are defined on different probability spaces, thus $V$ and $A$ are stochastically unrelated: there is no empirical context (e.g., trial number) in which a unimodal event $\{V\le s\}$ co-occurs with a unimodal event $\{A\le t\}$ to define a joint distribution for $(V,A)$. Nevertheless, such a joint distribution can always be constructed by the stochastic concept of \textit{coupling}. A \textit{coupling} of random variables $V$ and $A$ is a pair of random variables $(\hat{V},\hat{A})$ with a bivariate distribution function $H_{VA}(s,t)$ such that its marginal distributions are identical to $F_V$ and $F_A$ respectively, i.e.,
 \[V\overset{d}{=} \hat{V} \mbox{ and } A\overset{d}{=} \hat{A},\] 
where $\overset{d}{=}$ means ``equality-in-distribution''. Thus, existence of a coupling is equivalent to the assumption of ``context independence'' mentioned above. Inequality~(\ref{ind}) corresponds to an \textit{independent coupling} of $V$ and $A$ with 
\[  H_{VA}(s,t)= F_V(s)*F_A(t),\]
but there exists an infinite number of possible couplings\footnote{For a comprehensive treatment of the theory of coupling, see \cite{thorisson2000}.}. The ``trick'' is to find a dependence structure that fits one's purposes. 

For the race model Inequality~(\ref{rmi}), which can be written equivalently as
 \[F_{VA}(t) \le \min\{F_V(t)+F_A(t),1\}, \;t\ge 0,\]
it turns out that the right-hand side corresponds to the coupling of $V$ and $A$ generating maximal negative stochastic dependence between the two random variables. Moreover, the area $S$ between $F_{VA}$ and $\min\{F_V(t)+F_A(t),1\}$ equals the expected value of random variable $\min\{V,A\}$, i.e.,
\[ S = \mathrm{E}^-\min\{V,A\}, \]
under maximal negative dependence between $V$ and $A$, with superscript ``--'' indicating maximal negative dependence. 
\subsection*{CRE of RT under maximal negative dependence}
A measure of crossmodal response enhancement for reaction times, based on maximal negative dependence, can then be defined by replacing $\min\{\mathrm{E}RT_V,\mathrm{E}RT_A\}$ in Equation~\ref{crert}
by area $S$, yielding:
\begin{equation}\label{crert-max}
\mathrm{CRE_{RT}^-}=\frac{\mathrm{E}^-\min\{V,A\}-\mathrm{E}RT_{VA}}{\mathrm{E}^-\min\{V,A\}}\times 100.
\end{equation}
Because $\mathrm{E}^-\min\{V,A\}\le \min\{\mathrm{E}RT_V,\mathrm{E}RT_A\}$, it follows that
\[\mathrm{CRE_{RT}^-}\le \mathrm{CRE_{RT}}  \] always. In other words, the new index of crossmodal response enhancement for RT is more conservative than the traditional one. Proof of these statements, being analogous to the one given for spike counts in the next section, is omitted here, but see\cite{miller86,colonius1990possibly,Colonius2016}.

\section*{Measuring crossmodal enhancement in single neurons}
To fix ideas, let $N_V$, $N_A$, and $N_{VA}$ denote the random number of impulses (spikes) emitted in a given time interval by a neuron, following unisensory (visual, auditory) and crossmodal (visual-auditory) stimulation, respectively, without assuming any specific parametric distribution for these random variables. Inserting their expected values into the traditional $\mathrm{CRE}$ of Equation~(\ref{cre}) yields
\begin{equation}\label{msiva}
\mathrm{CRE_{SP}}=\frac{\mathrm{E}N_{VA}-\max\{\mathrm{E}N_V,\mathrm{E}N_A\}}{\max\{\mathrm{E}N_V,\mathrm{E}N_A\}}\times 100,
\end{equation}
where subscript $\mathrm{SP}$ indicates measurement of spikes. At the level of samples, the expected values are replaced by arithmetic averages. 

Realizations of random variables $N_V$ and $N_{A}$, with distribution functions $G_V$ and $G_A$, respectively, are collected across experimental trials under different stimulus conditions (unisensory and bisensory). Thus, as observed above for reaction times, they refer to distinct probability spaces and there is --a-priori-- no natural way to combine the results from unisensory visual and auditory trials. In particular, any assumption about stochastic (in-)dependence between $N_V$ and $N_{A}$ is void. Nevertheless, one can define a stochastic \textit{coupling} of the two random variables. Coupling of $N_V$ and $N_{A}$ here amounts to defining a  distribution $H_{VA}$ for a bivariate random vector  $(\tilde{N}_V,\tilde{N}_A)$ in such a way that its marginal distributions are identical to $G_V$ and $G_A$. 

Let $H_{VA}(m,n)=P(\tilde{N}_V\le m,\tilde{N}_A\le n)$, $m,n=0,1,\ldots$, be the distribution for some coupling of $N_V$ and $N_{A}$. As a bivariate (discrete) distribution, it obeys the Fr\'{e}chet inequalities valid for any distribution\cite{frechet51}:
\begin{equation}\label{frechet}
\max\{0,G_V(m) + G_A(n)-1\}\le H_{VA}(m,n) \le \min\{G_V(m),G_A(n)\},
\end{equation}
for all $m,n =0,1,\ldots$.  Setting $m=n$, we get
\[ H_{VA}(m,m)=P(\max\{\tilde{N}_V,\tilde{N}_A\}\le m), \]
and from (\ref{frechet}),
\begin{equation}\label{frechet2}
H^-(m)\equiv \max\{0,G_V(m) + G_A(m)-1\}\le H_{VA}(m,m) \le \min\{G_V(m),G_A(m)\}\equiv H^+(m),
\end{equation}
for $m=0,1,\ldots$. In (\ref{frechet2}) both upper bound $H^+(m)$ and lower bound $H^-(m)$ are univariate distribution functions of random variable $\max\{\tilde{N}_V,\tilde{N}_A\}$. Moreover, it is well known\cite{joe97} that $H^+$ and $H^-$ represent distributions with maximal positive, respectively negative, dependence between $\tilde{N}_V$ and $\tilde{N}_A$, assuming non-degenerate marginal distributions $G_V$ and $G_A$. 

\begin{proposition} 
Under any coupling of the univariate response random variables $N_V$ and $N_{A}$, the following bounds hold for expected value $\mathrm{E}\max\{N_V,N_A\}$,
\begin{equation*}\label{prop}
\max\{\mathrm{E}N_V, \mathrm{E} N_A\} \le\mathrm{E}\max\{N_V,N_A\}\le \mathrm{E}^- \max\{N_V,N_A\}, 
\end{equation*}
where $\mathrm{E}^- \max\{N_V,N_A\}$ is the expected value under maximal negative dependence between the univariate response random variables.
\end{proposition}
To prove the right-hand bound of the proposition, rewrite Equation~(\ref{frechet2}) as
\[1-H^+(m)\le 1-H_{VA}(m,m)=P(\max\{N_V,N_A\}>m) \le 1- H^-(m),\]
for $m=0,1,\ldots$.
Summing over all $m$ yields the result
\[
 \sum_{m=0}^\infty[1-H_{VA}(m,m)]=\mathrm{E} \max\{N_V,N_A\}\le \mathrm{E}^- \max\{N_V,N_A\}.
\]
The left-hand bound, $\max\{\mathrm{E}N_V, \mathrm{E} N_A\}\le \mathrm{E} \max\{N_V,N_A\}$ follows directly from \textit{Jensen's inequality} (see, e.g., \cite{ross96}  p. 51).
\subsection*{CRE in single neurons under maximal negative dependence}
From Proposition~1 it is clear that the sample value of $\mathrm{E}^- \max\{N_V,N_A\}$ is the largest mean obtainable from combining the unisensory responses via probability summation. Replacing $\max\{\mathrm{E}N_V, \mathrm{E} N_A\}$ by $\mathrm{E}^- \max\{N_V,N_A\}$ in the traditional $\mathrm{CRE}$ index results in the new index
\begin{equation}\label{msivanew}
\mathrm{CRE_{SP}^-}=\frac{\mathrm{E}N_{VA}-\mathrm{E}^- \max\{N_V,N_A\}}{\mathrm{E}^- \max\{N_V,N_A\}}\times 100.
\end{equation}
This new index measures the degree by which a neuron's observed multisensory response surpasses the level obtainable by optimally combining the unisensory responses (assuming that the neuron simply reacts to the more salient modality in any given cross-modal trial). The test for multisensory enhancement then amounts to comparing the observed mean number of impulses to crossmodal stimulation with the estimate for $\mathrm{E}^- \max\{N_V,N_A\}$. For empirical data, the expected value $\mathrm{E}N_{VA}$ is replaced by the sample mean of crossmodal responses and $\mathrm{E}^- \max\{N_V,N_A\}$ is estimated using the \textit{method of antithetic variates} as demonstrated below (see also \cite{ross96}). 

\subsubsection*{Two important consequences}Applying the new index has two important consequences. First, given that the new index is obviously always smaller or equal to the traditional index, \[\mathrm{CRE_{SP}^-}\le \mathrm{CRE_{SP}},\] some neurons previously labeled ``multisensory'' may lose that property under the new index. This is illustrated with an empirical data set following the next section.

Second, from the definition of $\mathrm{CRE_{SP}}$ it follows that changing the variability of the unisensory responses while leaving $\max\{\mathrm{E}N_V, \mathrm{E} N_A\}$ invariant, will not affect the value of $\mathrm{CRE}$. In contrast, the new index being based on $\mathrm{E}^- \max\{N_V,N_A\}$ can be sensitive to such changes. This is illustrated here for the case of Poisson distributed spikes. 
%
\subsection*{Example: Poisson-distributed spikes}
Let the spike counts $N_V$ and $N_A$ follow a Poisson distribution, i.e.,
\begin{equation}\label{poi}
P(N_i= m) = \exp[-\lambda_i] \frac{\lambda_i^m}{m!} \;\;\;\mbox{for $m=0,1,2\ldots$.}
\end{equation}
with $i=V$ or $i=A$. For this distribution, $\mathrm{E}N_i  =\lambda_i$ and, for the variance, $\mathrm{Var}N_i=\lambda_i$ as well. The traditional index can thus be written as
\begin{equation}\label{poi2}
\mathrm{CRE_{SP}}=\frac{\mathrm{E}N_{VA}-\max\{\mathrm{Var}N_V,\mathrm{Var}N_A\}}{\max\{\mathrm{Var}N_A,\mathrm{Var}N_A\}}\times 100,
\end{equation}
We assume, without loss of generality, that $\mathrm{Var}N_A < \mathrm{Var}N_V$.
Obviously, increasing $\mathrm{Var}N_A$ will not change the value of $\mathrm{CRE_{SP}}$ as long as  $\mathrm{Var}N_A$ is not strictly larger than $\mathrm{Var}N_V$. 
In contrast, as will now be shown, $\mathrm{E}^- \max\{N_V,N_A\}$, and therefore $\mathrm{CRE_{SP}^-}$ as well,  will not remain invariant with  $\mathrm{Var}N_A$ increasing. 

Inserting into the expected value yields
\begin{align}\label{poimean}
\mathrm{E}^-\max\{N_V,N_A\} &= \sum\limits_{m=0}^{\infty}[1-H^-(m)]=\sum\limits_{m=0}^{\infty}\left[1-\max\left\{0, \sum\limits_{k=0}^{m} P(N_V=k)+\sum\limits_{k=0}^{m} P(N_A=k)-1\right\}\right]\nonumber .
\end{align}
For given values of parameters $\lambda_V$ and $\lambda_A$, approximate computation of this expected value is simplified by using the fact\cite{Johnson1992} that the (cumulative) distribution for the Poisson is expressed in terms of the \textit{incomplete gamma function}. Specifically, for $i=V,A$:
\begin{equation*}
\sum\limits_{k=0}^{m} P(N_i=k)= \Gamma(m+1,\lambda_i) /\Gamma(m).
\end{equation*}
Here, the ratio $\Gamma(m+1,\lambda_i) /\Gamma(m)$ is the \textit{regularized incomplete gamma function} with $\Gamma(m)=(m-1)!$ and $\Gamma(m,\lambda_i)$ the \textit{incomplete gamma function}
\begin{equation}\label{gamma}
 \Gamma(m,\lambda_i)=  \int_{\lambda_i}^\infty e^{-t}t^{m-1}\,dt.
\end{equation}
For illustration of the effect, we choose specific, but otherwise arbitrary, parameter values: $\mathrm{E}N_{VA}=30$ and, for $\mathrm{Var}N_V=\lambda_V=22$ and $\mathrm{Var}N_V=\lambda_V=26$, we varied $\mathrm{Var}N_A=\lambda_A$  between $5$ and $22$ and $26$, respectively. Table~\ref{sc1-table} lists the corresponding values of $\mathrm{CRE_{SP}^-}$ as a function of $\mathrm{Var}N_V$ 
and $\mathrm{Var}N_A$ as well as the $\mathrm{CRE_{SP}}$ for the two different values of $\mathrm{Var}N_V$. Notably, increasing $\mathrm{Var}N_A=\lambda_A$ corresponds to a strong decrease in $\mathrm{CRE_{SP}^-}$, whereas $\mathrm{CRE_{SP}}$ remains invariant against such increase in variability of $N_A$. 
\begin{table}[ht]
\centering
\begin{tabular}{l|r|r|l}
$\mathbf{\lambda_V}$&$\mathbf{\lambda_A}$ & $\mathbf{{CRE_{SP}^-}}$&$\mathbf{{CRE_{SP}}}$\\
\hline
22&5&36.3&36.4\\
22&10&35.1 & \\
22&16&29.0 &\\
22&22&16.6 & \\
\hline
26&5&15.4&15.4\\
26&10&15.0 &\\
26&16&12.7 &\\
26&22&6.3 &\\
26&26&-0.2& 
\end{tabular}
\caption{\label{sc1-table} \textbf{Poisson-distributed spike counts.} Values of $\mathrm{CRE_{SP}^-}$ are shown as a function of $\lambda_A=\mathrm{Var}N_A$ and two fixed values of $\lambda_V=\mathrm{Var}N_V$. $\mathrm{CRE_{SP}^-}$ decreases with increasing variability of $N_A$, whereas $\mathrm{CRE_{SP}}$ remains constant. }
\end{table}
%
%
\section*{Empirical data}
First, we illustrate the computation of $\mathrm{CRE_{SP}^-}$ and $\mathrm{CRE_{SP}}$  for a single data set, recordings from a cat \textit{superior colliculus} (SC) neuron, followed by a comparison of both indexes on a larger number of such neurons. All data has been obtained from the lab of Mark T. Wallace \cite{Wallace}
\subsection*{Computing $\mathbf{CRE_{SP}^-}$ and $\mathbf{CRE_{SP}}$ for data from a single neuron}
The data set consists of the total number of spikes, recorded within a response window, that occurred from visual, auditory, and visual-auditory stimulation in $N=20$ trials, respectively (details in Table~\ref{sc2-table}). Spike numbers in the left-hand columns of Table~\ref{sc2-table} include spontaneous activity (S.A.), whereas the right-hand columns show the same recordings after S.A. was removed.

Note that a-priori there is no fixed correspondence between trial number and the individual values of $\mathbf{V}$ and $\mathbf{A}$. The antithetic variates method involves pairing the unisensory responses, sorted by increasing order ($\mathbf{V}$) and by decreasing order  ($\mathbf{A}$), and computing $\mathbf{max(V,A)}$ for each pair. Their mean value represents an estimate of $\mathrm{E}^- \max\{N_V,N_A\}$, that is, of the maximum expected value from combining the unisensory responses achievable via negatively dependent probability summation. The trial numbering of the $\mathbf{VA}$ values remains arbitrary.

\begin{table}[ht]
\centering

\begin{tabular}{r|cccc|cccr} 
	               &         \multicolumn{4}{c|}{\textbf{Spike numbers}}         &        \multicolumn{4}{c}{\textbf{Spike numbers w/o S.A.}}         \\
	               &              &              &                     &               &              &              &                     &  \\
	\textbf{trial} & $\mathbf{V}$ & $\mathbf{A}$ & $\mathbf{max(V,A)}$ & $\mathbf{VA}$ & $\mathbf{V}$ & $\mathbf{A}$ & $\mathbf{max(V,A)}$ & $\mathbf{VA}$ \\
	\hline
	             1 &      3       &      8       &          8          &      11       &    1.113     &    7.493     &        7.493        &        18.933 \\
	             2 &      4       &      8       &          8          &      22       &    2.113     &    7.493     &        7.493        &        13.933 \\
	             3 &      5       &      7       &          7          &      17       &    3.113     &    6.493     &        6.493        &        15.933 \\
	             4 &      5       &      7       &          7          &      19       &    3.113     &    6.493     &        6.493        &        14.933 \\
	             5 &      5       &      7       &          7          &      18       &    3.113     &    6.493     &        6.493        &         9.933 \\
	             6 &      6       &      7       &          7          &      13       &    4.113     &    6.493     &        6.493        &        14.933 \\
	             7 &      6       &      6       &          6          &      18       &    4.113     &    5.493     &        5.493        &         7.933 \\
	             8 &      7       &      6       &          7          &      11       &    5.113     &    5.493     &        5.493        &        22.933 \\
	             9 &      7       &      6       &          7          &      26       &    5.113     &    5.493     &        5.493        &        16.933 \\
	            10 &      8       &      6       &          8          &      20       &    6.113     &    5.493     &        6.113        &        24.933 \\
	            11 &      8       &      6       &          8          &      28       &    6.113     &    5.493     &        6.113        &        15.933 \\
	            12 &      9       &      6       &          9          &      19       &    7.113     &    5.493     &        7.113        &        21.933 \\
	            13 &      9       &      5       &          9          &      25       &    7.113     &    4.493     &        7.113        &        11.933 \\
	            14 &      10      &      5       &         10          &      15       &    8.113     &    4.493     &        8.113        &        13.933 \\
	            15 &      10      &      5       &         10          &      17       &    8.113     &    4.493     &        8.113        &        15.933 \\
	            16 &      10      &      4       &         10          &      19       &    8.113     &    3.493     &        8.113        &        15.933 \\
	            17 &      11      &      4       &         11          &      19       &    9.113     &    3.493     &        9.113        &        14.933 \\
	            18 &      11      &      4       &         11          &      18       &    9.113     &    3.493     &        9.113        &        27.933 \\
	            19 &      13      &      4     &         13          &      31       &    11.113    &    3.493     &       11.113        &        13.933 \\
	            20 &      14      &      4       &         14          &      17       &    12.113    &    3.493     &       12.113        &         7.933 \\
	\hline 
	 mean &     8.05     &     5.75     &        8.85         &     19.15     &    6.163     &    5.243     &        7.484        &        16.083 \\
	   standard dev. &    2.999     &    1.333     &        2.159        &     5.204     &    2.999     &    1.333     &        1.791        &         5.204 \\ 
\end{tabular} 
\caption{{Sample of recordings from a single cat SC}:  Columns 2 and 6 ($\mathbf{V}$) are arranged by increasing order, 3 and 7 ($\mathbf{A}$)  by decreasing order. S.A. stands for ``spontaneous activity'' ($4.26$ spikes/s  in this sample). Standard PSTHs were computed. Spontaneous activity was computed from the $500$ ms preceding each stimulus onset (allowing at least $1500$ ms between each trial). A threshold of mean S.A. rate per $10$ ms bin plus 2 standard deviations was computed, only used to determine onset and offset. Response onset was defined when the first spike occurred within the bin that rises above this threshold and remained above for at least 3 bins. Offset was counted as the last spike in the bin just before the response fell back below this threshold and remained below for 3 bins. The response window (duration) is the time between onset and offset. Total number of spikes (left columns in the table) include all spikes within the response window, which will inevitably include some S.A. The right columns include responses with S.A. removed. The expected number of S.A. spikes within the given window (i.e., S.A. times window size in seconds) was removed. This is never an integer and can sometimes cause negative values on some trials. This number represents ``change from baseline firing'' (information obtained from M. T. Wallace, personal communication, July 18, 2015)\label{sc2-table}}
\end{table}

Computing the traditional $\mathrm{CRE_{SP}}$ value by inserting the estimates from Table~\ref{sc2-table} in Equation~\ref{msiva} , i.e., replacing the expected values by the means,  yields

\[ \mathrm{CRE_{SP}}=\frac{\mathrm{E}N_{VA}-\max\{\mathrm{E}N_V,\mathrm{E}N_A\}}{\max\{\mathrm{E}N_V,\mathrm{E}N_A\}}\times 100 \approx \frac{19.15-\max\{8.05,5.75\}}{\max\{8.05,5.75\}}\times 100=137.89 [\%].\]
for spike numbers containing S.A (left-hand columns). The corresponding value for the new index is estimated by inserting the estimates from Table~\ref{sc2-table} in Equation~\ref{msivanew},
\[
\mathrm{CRE_{SP}^-}=\frac{\mathrm{E}N_{VA}-\mathrm{E}^- \max\{N_V,N_A\}}{\mathrm{E}^- \max\{N_V,N_A\}}\times 100 \approx \frac{19.15-8.85}{8.85}\times 100=116.64 [\%].\]
The corresponding values for responses with S.A. removed (right-hand columns) amount to 
\[ \mathrm{CRE_{SP}} \approx \frac{16.083-\max\{6.163,5.243\}}{\max\{6.163,5.243\}}\times 100=160.96 [\%].\]
and
\[
\mathrm{CRE_{SP}^-}\approx \frac{16.083-7.484}{7.484}\times 100= 114.90 [\%].\]
The results are quite clearcut. For this neuron, replacing $\mathrm{CRE_{SP}}$ by $\mathrm{CRE_{SP}^-}$ corresponds to a drop from about $161 \%$ to about $115 \%$ with spontaneous activity removed, and from about $138 \%$ to about $117 \%$ when spontaneous activity was retained. Thus, applying the new index may well lead to dropping the ``multisensory'' label for this neuron depending, of course, on one's criterion for attaching that label.
\subsection*{Comparing $\mathbf{CRE_{SP}^-}$ and $\mathbf{CRE_{SP}}$ for $n=27$ recording blocks}
The total data set comprised 84 recording blocks from 20 SC cells of length 15 each, where the number of spikes to visual-auditory stimulation was found significantly larger than the maximum of responses to unisensory stimulation, according to the categorization from the Wallace lab. In 57 of these blocks, there was no response at all from either visual or auditory stimulation. For those cases, $\mathrm{CRE_{SP}^-}=\mathrm{CRE_{SP}}$ by definition, so comparison is void. The data from the remaining 27 recording blocks were available for comparing both indexes.  

\begin{figure}
\begin{center}
\includegraphics[scale=.65]{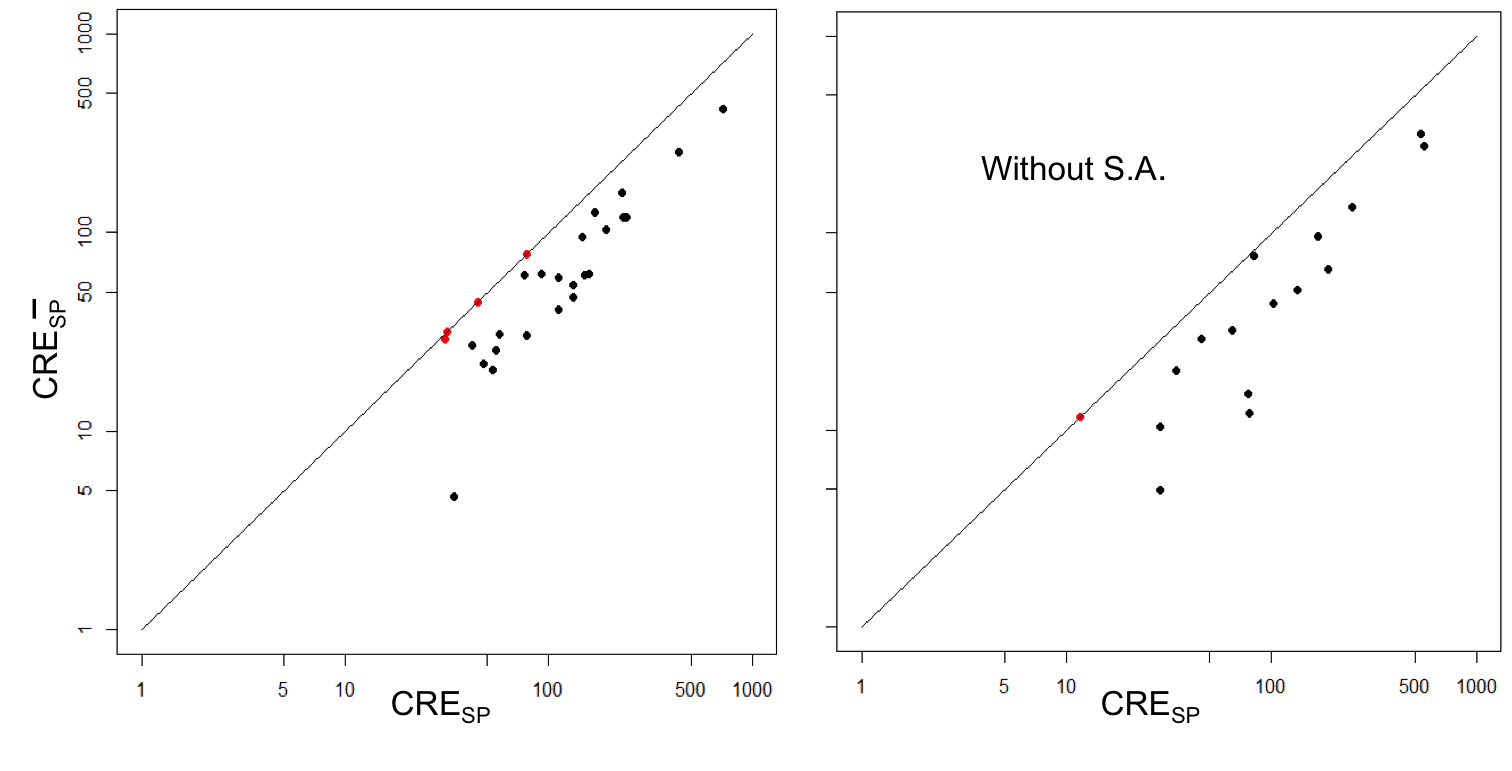}
\end{center}
\caption{Bootstrapped values (logarithmic scale) of $\mathrm{CRE_{SP}^-}$ vs. $\mathrm{CRE_{SP}}$ based on 10,000 samples (in right-hand panel, spontaneous activity was removed). Except for 4 out of 27 cases (left panel) and 1 out of 19 cases (red points), the new index was significantly smaller than the traditional one ($95 \%$ confidence intervals, too small to be shown on the graphs). }
\label{fig:cre}
\end{figure}

In order to obtain confidence interval estimates for the difference between $\mathrm{CRE_{SP}^-}$ and $\mathrm{CRE_{SP}}$, each of these 27 blocks underwent a bootstrap procedure, i.e., 10,000 random samples of $N=15$ were taken with replacement from the sets of spike frequencies for visual ($\mathbf{V}$), auditory ($\mathbf{A}$), and bimodal ($\mathbf{VA}$) stimulation. For each sample, both $\mathrm{CRE_{SP}^-}$ and $\mathrm{CRE_{SP}}$ were computed yielding a $95 \%$ confidence interval for their difference in each of the 27 recording blocks. The points of Figure \ref{fig:cre} depict pairs of bootstrap estimates of $(\mathrm{CRE_{SP}},\mathrm{CRE_{SP}^-})$. In the left panel (with spontaneous activity retained), there were 4 out of 27 cases with no significant difference between both measures (red color), after spontaneous activity was removed, only 1 out of 19 cases was not significant (see right panel). In the latter, the number of possible comparisons decreased to 19 because in the other blocks there was no activity left for one of the unisensory conditions.
In summary, this arguably limited data set supports the observation that many neurons previously labeled ``multisensory'' will no longer be categorized as such.
\section*{Discussion and Conclusion}
The issue of how to quantify crossmodal response enhancement due to the occurrence of multisensory integration has been under discussion in both behavioral and neurophysiological research. The most widely used index up to now expresses crossmodal enhancement as a proportion of the strongest unisensory response. It has descriptive value but lacks a theoretical basis. Such a foundation is essential because, as widely acknowledged in both reaction time and neural studies, being responsive to multiple sensory modalities does not guarantee that the response has been generated by actually integrating the multiple sensory inputs, rather than simply responding to the most salient stimulus modality. Here we suggest a new index that measures by how much the crossmodal response surpasses the level obtainable by optimally combining the unisensory responses. Optimality is achieved by using a probability summation mechanism that combines the unisensory responses with maximal negative dependence. Importantly, no claim is made that the system actually operates under this mechanism, it only serves as well-defined benchmark against which to gauge the crossmodal response. 

It has been demonstrated here that the new index can be defined in a consistent manner for both studying reaction times and responses by single neurons (spike frequencies). Whereas the index is closely linked to the race model inequality, a widely used testing procedure for multisensory integration in reaction times, its application to neural responses has new and potentially important consequences: neurons previously labeled as ``multisensory'' may lose that property since the new index tends to yield smaller values for the amount of crossmodal enhancement. This was exemplified here with a data set collected from single SC neurons. The extent to which this holds more generally can only be determined by a large-scale investigation of a multitude of neurons from empirical studies. Obviously, at the level of a (sub-)population of neurons, such a relabeling may lead to a reassessment of the distribution of multisensory neurons and different types of unisensory neurons for that region. Moreover, studies probing the entire scope of the behavior of multisensory neurons, e.g. by looking at intrinsic differences in the dynamic range of these neurons (see \cite{perrault2005}), may come to different conclusion when using the new index.

We also showed that the new index, $\mathrm{CRE_{SP}^{(-)}}$,  is easy to compute and does not require any specific assumption about the distribution of spikes. The special case of Poisson-distributed spikes was drawn upon to demonstrate that the new index is sensitive to the variability in the data, in contrast to the traditional index which by definition only depends on the means of the uni- and crossmodal response distributions. 

It is worth mentioning that the new approach can also be applied to an alternative measure, comparing cross-modal responses to the sum of the unisensory responses (``additive model'') (see also \cite{stanford2005})~. From \cite{rueschendorf82} (and more recent papers in actuarial statistics), it is possible to compute the maximally achievable sum of two random variables and, using the same logic as for computing $\mathrm{CRE_{SP}^{(-)}}$, cross-modal responses can be compared with the response level obtainable by adding the unisensory responses in an optimal way.
%
%

Future research should address a number of issues. For example, is the new index consistent with the ``inverse effectiveness rule'' of multisensory integration, stating that crossmodal response enhancement decreases with the intensity of the stimuli applied ? Preliminary reasoning suggests that both $\mathrm{CRE_{SP}}$ and $\mathrm{CRE_{SP}^-}$ are consistent with this ``rule''. Only the latter, however, seems also sensitive to an increase in the intensity of the modality to which the neuron is less responsive.

Another issue is whether the logic of the new index can be extended to more than two modalities? Such a generalization is not straightforward given that maximal negative dependence among three random variables is strongly limited. On a broader level, it would be interesting to explore whether the new index, or at least its logic, could be utilized beyond the level of single neuron responses, possibly including data from functional magnetic resonance studies\cite{Klemen2012}. As the authors of a recent review\cite{Goebel2009} put it, `` ..., an enhanced BOLD response for multisensory relative to unisensory stimulation can be due to ``true'' multisensory neurons integrating stimulation from two or more sensory modalities, but it can just as well be explained by driving two unisensory sub-populations instead of one. If the latter scenario would be true, one might wrongly infer multisensory integration at the neuronal level.''

Given the recent results by Miller et al.\cite{miller2015}, showing ``...that the integration of temporally displaced sensory responses is also highly dependent on the relative efficacies with which they drive their common target neuron'',  one may also more generally question the usefulness of any static measure of crossmodal enhancement, and this may lead to implementing a temporal dimension to a quantitative index of crossmodal enhancement.

\bibliography{colonius_refs}

\section*{Acknowledgements}
We are most grateful to Mark Wallace, Aaron Nidiffer, and collaborators (Vanderbilt University) who kindly made available their data set. 
 
\section*{Author contributions statement}


%
%

%
\end{document}